\documentclass[twocolumn,showpacs,pre,aps]{revtex4}

\usepackage{graphicx}
\usepackage{dcolumn}
\usepackage{bm}

\begin{document}

\title{On In-Plane Current Distribution Producing a Given
Axisymmetric \\ Distribution of Normal-to-Plane Magnetic Field}

\author{Yu.\,E.\,Kuzovlev}
\email{kuzovlev@kinetic.ac.donetsk.ua}
\affiliation{Donetsk
Physics and Technology Institute, 83114 Donetsk, Ukraine}


\begin{abstract}
A very good as well as simple approximate solution is suggested to
the problem which flat circular distribution of electric current in
finite film produces a given axially symmetric distribution of
normal-to-film magnetic field.
\end{abstract}


\maketitle

1. In films of type II superconductors, a concentration, $\,n\,$, of
vortices piercing the film usually simply correlates with magnetic
flux density, $\,\bm{B}\,$ , at its surface:  $\,\Phi_0 n =
(\bm{B}-\bm{B}_0)_n\,$, where $\,(…)_n\,$ means normal-to-film
component of a vector,  $\,\Phi_0\,$  is magnetic flux quantum,
$\,\bm{B}_0\,$  - uniform external field, and $\,\bm{B}-\bm{B}_0\,$
is magnetic field induced by (super-)current flowing in the film,
$\,\bm{J}\,$. The same current stimulates the vortices to move. To
consider this motion, analytically or numerically, we should be able
to determine what is distribution of $\,\bm{J}\,$ which produces a
given distribution of $\,\bm{B}-\bm{B}_0=\Phi_0 n\,$.

Such task is non-trivial even under flat geometry, if the film
physically occupies a part of its plane and correspondingly the field
$\,\bm{B}\,$  is considered to be known in that part only.
Previously, an exact analytical solution of the task was found for
the single case of infinitely long flat strip \cite{chg,b}, assuming
homogeneity of $\,\bm{B}\,$ along it.

Recently, I suggested exact solution for one more case of round film
\cite{i}, assuming axial symmetry of $\,\bm{B}\,$. The obtained
radial distribution of circular current, $\,J(r)\,$, as functional of
radial distribution of $\,H(r)\equiv (\bm{B}-\bm{B}_0)_n\,$, can be
represented by sum of regular term and singular term:
\begin{equation}
\begin{array}{c}
J(r) = J_{REG}(r) + J_{SING}(r)\,\,\,,\label{sc}
\end{array}
\end{equation}
\begin{equation}
J_{REG} (r) =  - \int_0^1 \,T(r,\rho )\rho\,\frac{dH(\rho )}{d\rho }
\,d\rho \,\,,\label{r}
\end{equation}
\begin{equation}
J_{SING} (r) = -\frac{2H_e\,r}{\pi\sqrt {1-r^2}}\,\,,\label{s}
\end{equation}
\begin{equation}
H_e\equiv -\int_0^1 \frac{H(\rho )\rho \,d\rho }{\sqrt {1-\rho ^2
 }} \,\label{he}
\end{equation}
Here, $\,J\,$ has the sense of the current density, $\,j\,$, multiplied
by $\,2\pi/c\,$ (in CGS units) and integrated over film's thickness;
 film's radius is taken to be unit of length;  the kernel
$\,T(r,\rho)\,$ is defined by formulas
\begin{equation}
T(r,\rho ) = \frac{2}{\pi }\frac{K(k) - F(\varphi ,k) - E(k) +
E(\varphi ,k)}{\min (r,\rho )}\,\,,\label{t}
\end{equation}
\[
k = \frac{\min (r,\rho)}{\max (r,\rho )}\,\,, \,\,\,\varphi = \arcsin
\,\max \,(r,\rho )\,\,,
\]
where standard designations of the complete and incomplete elliptic
integrals of 1-st and 2-nd kinds are used (see e.g. \cite{kk}). I
permit myself to copy from [3] the plots of $\,T(r,\rho)\rho\,$ (see
Fig.1 below).

Notice, first, that the singular contribution $\,J_{SING}(r)\,$ looks
exactly as Meissner current in vortex-free film under effective
uniform external field $\,H_e\,$. Second, the regular part of the
current, $\,J_{REG}(r)\,$, turns into zero (or at least remains
finite) at film's edge. Thus the condition $\,H_e=0\,$ ensures the
absence of any current singularity at the edge. At that, the relation
between $\,J(r)\,$ and $\,H(r)\,$ becomes non-local analogue of the
local one, $\,4\pi j/c = -dH/dr\,$, which takes place under cylindric
geometry.

2. In general, it may be convenient to unify $\,J_{REG}(r)\,$ and
$\,J_{SING}(r)\,$ into the single expression \cite{i}:
\begin{equation}
J(r) =  -\frac{d}{{dr}}\,\,\int_0^1 {G(r,\rho )\rho\,H(\rho )
\,\,d\rho }\,\,,\label{tr}
\end{equation}
\begin{equation}
G(r,\rho)\,=\,\frac {2}{\pi\,(\rho
+r)}\left[K(k)-2F(\phi,k)\right]\,\,,\label{G}
\end{equation}
\[
k=\frac {2\sqrt{\rho r}}{\rho +r}\,\,,\,\,\,\phi=\frac 12\,(\,\arcsin
\,\rho+\arcsin\,r\,)
\]

Both (\ref{t}) and (\ref{G}) contain incomplete elliptic integrals.
But, unfortunately, not all of popular mathematical programs can
quickly (or let somehow) calculate them. This fact just implies the
subject of present article, namely, approximation of the kernel
$\,G(r,\rho)\,$ in terms of complete elliptic integrals only.

For this purpose, we can apply the series expansion of the kernel
$\,G(r,\rho)\,$ (see \cite{i}):
\begin{equation}
G(r,\rho)\,\equiv \,\frac {2}{\pi
}\,\sum_{k=0}^{\infty}\,(4k+3)\,S_k^2\, P_{2k+1}\left
(r^{\prime}\right )P_{2k+1}\left (\rho^{\prime}\right )\,\,
\label{series}
\end{equation}
where\,\,\, $\,\,x^{\prime}\equiv
\sqrt{1-x^2}\,$\,,\,$\,\,\rho^{\prime}\equiv \sqrt{1-\rho^2}\,$\,,
\begin{equation}
\begin{array}{c}
S_k\,\equiv\, B(1/2,k+1)/2\,=\,2^k
k!/(2k+1)!!\,\,\,,\,\,\,\,\,\,\label{sk}
\end{array}
\end{equation}
with $\,B(x,y)=\Gamma(x)\Gamma(y)/\Gamma(x+y)\,$ being the
beta-function, and $\,P_m\,$ are ordinary Legendre polynomials
\cite{kk}. The key to the desired simplification is in the
inequalities
\begin{equation}
1\leq\, \frac {4k+3}{12}\,B^2\left(\frac 12,k+1\right)\,< \frac
{\pi}{3}\approx 1.0472 \,\,,\label{ap}
\end{equation}
which can be easy proved with the help of the Stirling formula.

\begin{figure}
\includegraphics{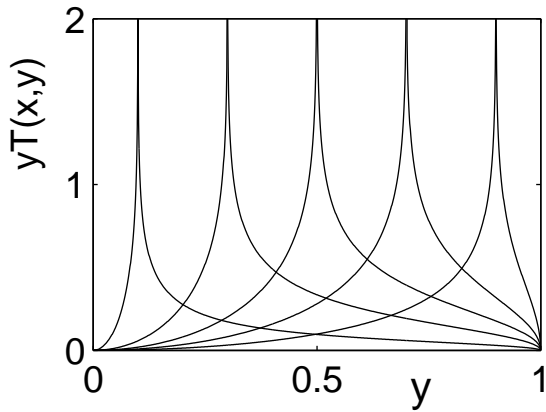}
\caption{\label{fig1} $\,T(x,y)\,y\,$ via $\,y\,$ at
$\,x\,=\,0.1,\,0.3,\,0.5,\,0.7$ and $0.9\,$. }
\end{figure}

\begin{figure}
\includegraphics{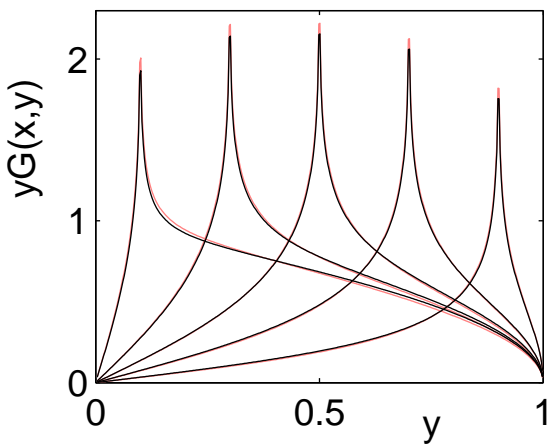}
\caption{\label{fig1} $\,G(x,y)\,y\,$ (pink curves) in comparison
with $\,G_{app}(x,y)\,y\,$ (black curves) via $\,y\,$ at
$\,x\,=\,0.1,\,0.3,\,0.5,\,0.7$ and $0.9\,$. }
\end{figure}

Our approximation will be merely replacement of the factor $\,
(4k+3)S_k^2\,$ by its lower bound, that is by $\,3\,$, in accordance
with (\ref{ap}) and (\ref{sk}). Corresponding approximate series for
the kernel $\,G(r,\rho)\,$ is
\begin{equation}
G(r,\rho)\,\approx G_{app}(r,\rho)= \,\frac {6}{\pi
}\,\sum_{k=0}^{\infty}\,P_{2k+1}\left (r^{\prime}\right
)P_{2k+1}\left (\rho^{\prime}\right )\,\, \label{apser}
\end{equation}
Fortunately, its sum can be fully expressed via complete elliptic
integrals, if use the representation
\[
\sum_{k=0}^{\infty}\,P_{2k+1}(x)P_{2k+1}(y)\,=\,\frac 12\,
[F(x,y)-F(x,-y)]\,\,,
\]
\[
F(x,y)\,\equiv \int_{-\pi}^{\pi}\,P\left(e^{i\phi},x\right)
P\left(e^{-i\phi},y\right)\,\frac {d\phi}{2\pi}\,\,\,,
\]
where $\,P(t,u)\,$ is generating function of the Legendre
polynomials,
\[
P(t,u)\,=\sum_{n=0}^{\infty}\,t^nP_n(u)=\frac
{1}{\sqrt{1+t^2-2tu}}\,\,\,,
\]
and then reduce $\,F(x,y)\,$ to standard elliptic-type integrals
\cite{kk,dw}. The result is
\begin{equation}
G_{app}(x,y)\,=\,\frac{3\sqrt{2}}{\pi
^2}\left[\frac{K(k_{-})}{\sqrt{S_{-}}} -\frac{K(k_{+})}{\sqrt{S_{+}}
}\right]\,\,\,,\label{appg}
\end{equation}
\begin{equation}
S_\pm \equiv 1 + xy \pm \sqrt {(1 - x^2 )(1 - y^2 )}\,\,\,,
\,\,\,\,k_\pm \equiv \frac{2xy}{S_\pm} \label{ks}
\end{equation}

Fig.2 shows this approximation in comparison with formally exact
kernel (in your mind please continue the logarithnic peaks to 
infinity). In practice, relative error of the approximation does 
not exceed 3\,\%\,.

I am grateful to Dr.~Yu.\,Genenko and Dr.~S.\,Yampolskii for useful
comments and attracting my attention to the work \cite{b}.

\end{document}